\documentclass[electronics,article,accept ,moreauthors,pdftex]{mdpi}

\firstpage{1} 
\pubvolume{9}
\issuenum{6}
\articlenumber{914}
\pubyear{2020}
\copyrightyear{2020}
\history{Received: 3 May 2020; Accepted: 28 May 2020; Published: 30 May 2020}

\usepackage{url}
\usepackage{multirow}
\usepackage[ruled,vlined]{algorithm2e}
\usepackage{subcaption}

\Title{Blended Multi-modal Deep ConvNet Features for Diabetic Retinopathy Severity Prediction}

\Author{Jyostna Devi Bodapati $^{1}$, N Veeranjaneyulu $^{2}$, Shaik Nagur Shareef$^{1}$, Saqib Hakak $^{3}$, Muhammad Bilal$^{4}$, Praveen Kumar Reddy M $^{5}$, Ohyun Jo $^{6*}$}

\AuthorNames{Firstname Lastname, Firstname Lastname and Firstname Lastname}

\address{%
	$^{1}$ \quad Department of CSE, Vignan's Foundation for Science Technology and Research, Guntur-522213, India \\
	$^{2}$ \quad Department of IT, Vignan's Foundation for Science Technology and Research, Guntur-522213, India \\
	$^{3}$ \quad Faculty of Computer Science, University of Northern British Columbia, Canada,
	\\
	$^{4}$ \quad Dept. of Computer and Electronics Systems Engineering, Hankuk University of Foreign Studies, Yongin-si, 17035, Korea \\
	$^{5}$ \quad School of Information Technology and Engineering, VIT, Vellore, 632014, India\\
	$^{6}$ \quad Department of Computer Science, College of Electrical and Computer Engineering, Chungbuk National University, Cheongju 28644, South Korea\\
}

\corres{Correspondence: ohyunjo@chungbuk.ac.kr}

\abstract{Diabetic Retinopathy (DR) is one of the major causes of visual impairment and blindness across the world. It is usually found in patients who suffer from diabetes for a long period. Major focus of this work is to derive optimal representation of retinal images that further helps to improve the performance of DR recognition models. In order to extract optimal representation, features extracted from multiple pre-trained ConvNet models are blended using proposed multi-modal fusion module. These final representations are used to train a Deep Neural Network (DNN) used for DR identification and severity level prediction. As each ConvNet extract different features, fusing them using 1-D pooling, and cross pooling lead to better representation than using features extracted from a single ConvNet. Experimental studies on benchmark Kaggle APTOS 2019 contest dataset reveals that the model trained on proposed blended feature representations is superior to the existing methods. In addition, we notice that cross average pooling based fusion of features from Xception and VGG16 is the most appropriate for DR recognition. With the proposed model, we achieve an accuracy of 97.41\%, and a kappa statistic of 94.82 for DR identification and an accuracy of 81.7\% and a kappa statistic of  71.1\% for severity level prediction. Another interesting observation is that, DNN with dropout at input layer converges faster when trained using blended features, than compared to the same model trained using uni-modal deep features.}


\begin{document}


\section{Introduction}
\begin{figure*}[t!]
    \centering
    \label{DR_image}
    \begin{subfigure}[t]{0.5\textwidth}
        \centering
        \includegraphics[width=6cm]{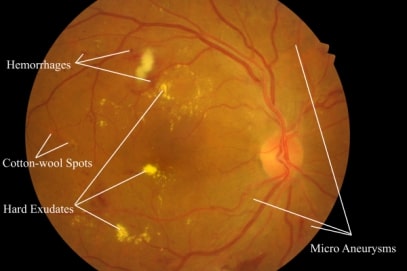}
        \caption{}
        \label{fig:DR}
    \end{subfigure}
    \begin{subfigure}[t]{0.5\textwidth}
        \centering     \includegraphics[width=6cm]{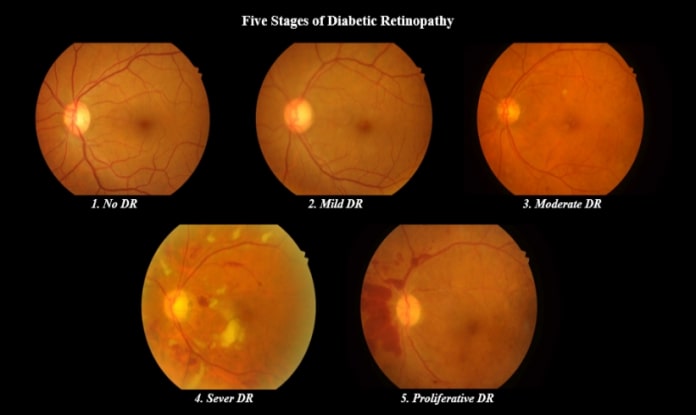}     \caption{}
        \label{fig:nodr}
    \end{subfigure}
    \caption{Samples of DR effected fundus images: (a)  Types of lesions formed (b) levels of severity}
    \label{fig:dr_severity}
\end{figure*}

Diabetic Retinopathy (DR) is an adverse effect of Diabetes Mellitus(DM) \cite{cheung2008retinal} that leads to permanent blindness in humans. It is usually caused by the damage to blood vessels that provide nourishment to light-sensitive tissue called the retina. As per statistics \cite{flaxmanglobal}, DR is the fifth leading cause for blindness across the globe. According to the World Health Organization (WHO), by 2013, around 382 million people are  suffering from DR, and this number may rise to 592 million by 2025. It is possible to save many people from going blind if DR is identified in the early stages. Small lesions are formed in the eyes of DR effected people and the type of lesions formed decides the level of severity of DR. Figure \ref{fig:DR} shows  types of lesions that include Micro Aneurysms (MA), Exudates, Hemorrhages, Cotton Wool Spots and improperly grown blood vessels on the retina. 

DR can be categorised into five different stages \cite{gulshan2016development}:  
No DR (Class-0), Mild DR (Class-1), Moderate DR (Class-2), Severe DR (Class-3) and Proliferative DR (Class-4). Sample retinal images with different severity levels of DR are shown in the figure \ref{fig:nodr}. Mild DR is the early stage during which the formation of Micro Aneurysms (MA) can be observed. As the disease progresses to Moderate stage,  swelling of blood vessels can be found, which leads to blurred vision. During the later  Non-Proliferative DR (NPDR) stage, abnormal growth of blood vessels can be noticed. This stage is severe due to the blockage of a large number of blood vessels. Proliferative DR (PDR) is the advanced stage of DR, during this stage retinal detachment along with large retinal break can be observed that leads to complete vision loss \cite{williams2004epidemiology}. 

In traditional DR diagnosis approaches, manual grading of the retinal scan is required to identify the presence or absence of retinopathy. If DR is confirmed as Positive, further diagnosis is recommended to identify severity level of the disease. This kind of diagnosis is quiet expensive and time consuming as it demands human expertise. If DR identification is automated then diagnosis of the disease becomes affordable to many people. In the recent past, several machine learning tools have been introduced to address the same.

Early approaches to DR identification,  where the presence or absence of DR is revealed, focuses on spotting the Hard Exudates (HEs). A dynamic threshold based Support Vector Machine (SVM) is used to segment HE in the retinal images \cite{long2019automatic}. Fuzzy C-means is used to detect HE and SVM is used to identify severity level of the disease to make the system more sophisticated \cite{haloi2015gaussian}. SVM based classifiers are adapted to find cotton wool spots in the retinal images.

With the introduction of deep learning, focus of the researchers has been shifted from spotting HEs to MAs. A two step CNN is introduced to segment MAs in the given retinal scans \cite{noushin2019microaneurysm}. Another CNN architecture that is trained using selective sampling approach is proposed to detect hemorrhages \cite{grinsven2016fast}. A max-out activation is introduced to improve the performance of a DNN model for which DR is used as an application to find MA \cite{haloi2015improved}. Recently a  bounding box based approach is introduced to identify the region of interest in the retinal images    \cite{srivastava2017detecting}. Though good number of methods are available in the literature, they are either sub-optimal or complex. Hence there is a need for a solution that is simple and robust. 

The objective of this work is to design a simple and robust deep learning-based approach to recognize DR from the given retinal images.
Major focus this work is to obtain a better feature representation of the retinal images which ultimately leads to the better model and to accomplish this, we propose  Uni-modal and Multi-modal approaches. Initially, for the given retinal images, deep features are extracted from different pre-trained ConvNets like VGG16, NASNet, Xception and Inception ResNetV2. In Uni-modal approach, features  extracted from a single pre-trained ConvNet gives the final feature representation. In multi-modal approach, our idea is to blend the deep features extracted from multiple ConvNets to get  the final feature representation. We propose different pooling based approaches to blend multiple deep features. 
To check the efficiency of our feature representation, a Deep Neural Network (DNN) architecture is proposed for identification of DR (task1) and to recognize severity level of DR (task2). We observe that in multi-modal approach, blending deep  features from Xception and Inception ResNet V2 outperforms others in both the tasks. Another interesting observation is that there is a drop in the number of false positives which is most desirable. Experimental studies on the benchmark APTOS 2019 dataset reveals that our blended feature representations trained using DNN model gives a superior performance compared to the existing methods.

Following are the major contributions of the proposed work:
\begin{itemize}
\item Effectiveness of the uni-modal feature representation is verified. 
\item A blended multi-modal feature representation approach is introduced 
\item Different pool based approaches are proposed to blend deep features.
\item A DNN architecture with dropout at the input layer is proposed to test the efficiency of the proposed uni-modal and blended multi-modal feature representations.
\item APTOS 2019 benchmark dataset is used to compare the performance of the proposed approach with existing models 
\end{itemize}

\section{Related Work}

In the resent past, machine learning models are very popular to solve various problems like image classification \cite{bodapati2019feature}, text processing \cite{bodapati2019sentiment}, real-time fault diagnosis \cite{zhuo2018real} and healthcare \cite{xia2020ensemble, moreira2019comprehensive}. It is very common to use ML algorithms to address disease prediction \cite{gadekallu2020deep,patel2020review} \cite{reddy2019hybrid}. 

In this section we report various conventional models available in the literature for the task of DR recognition. In \cite{wu2013classification}, an easy to remember scientific approach has been introduced for DR severity identification. In \cite{akram2013identification}, the authors presented a hybrid classifier by using both GMM and SVM as an ensemble model to improve the accuracy of the model. The same approach has been modified by augmenting the feature set with shape, intensity, and statistics of the affected region \cite{akram2014detection}.  A random forest-based approach is proposed in \cite{casanova2014application} \cite{verma2011detection} and   segmentation based approaches are proposed in \cite{welikala2014automated}. In \cite{welikala2015genetic}, a genetic algorithm-based feature extraction method is introduced. Different shallow Classifiers such as the GMM, KNN, SVM, and AdaBoost are being analysed \cite{roychowdhury2013dream} to differentiate lesions from non-lesions. A hybrid feature extraction based approach is used in \cite{mookiah2013evolutionary}.

In the next few lines, deep learning models available in the literature for the task of DR severity identification are introduced. A large dataset consisting of 1,28,175 retinal images is used and trained using deep CNN. In \cite{porter2019whole} data augmentation method is used to generate the data on CNN architecture. Fuzzy models are used in \cite{rahim2016automatic}, a hybrid model that is designed based on fuzzy logic, Hough Transform and numerous extraction methods are being implemented as part of their system. A combination of fuzzy C-means and deep CNN architectures are used in \cite{dutta2018classification}. A Siamese Convolutional Neural Network is used in \cite{zeng2019automated} to detect diabetic retinopathy.      

With the introduction of deep learning models, focus has been shifted to deep feature based models. In \cite{mateen2019fundus} Muhammad Mateen used features extracted from different layers of pre-trained ConvNet like VGG19 and further applied PCA and SVD on those features, for dimension reduction \cite{9036908} to avoid over-fitting. In the case of former models, the model is not robust, and in the latter case, the models are robust, but large datasets are needed to train the model. A PCA based fire-fly model \cite{bhattacharya2020novel} along with deep neural network is used for DR detection \cite{gadekallu2020early}, UCI repository is used for the experiments.

Performance of any ML algorithm is subject to the features extracted from the given data.  
Conventional ML models need a separate algorithm (GIST, HOG and SIFT) for feature learning and gives a global or local representation of the images  and the features.  Features extracted in this process are known as hand crafted features. Till the entry of deep learning models, these handcrafted features were dominant and being widely used for feature extraction.

\subsection{Deep ConvNets for feature extraction and transfer learning}
 Deep learning models\cite{jindal2018drums,vinayakumar2020visualized,alazab2020multidirectional} learn the essential characteristics of the input images.  This exceptional capability of the deep models make them representation models, as these models can represent the data efficiently and reduce the use of the additional feature extraction phase where features are handcrafted. Deeper layers of the CNN models can represent the entire given input efficiently than the early layers. 

The downside of the deep learning models is that they need enormous amounts of data for training, which is usually scarce for most of the real-time applications. This problem can be addressed by the introduction of transfer learning, where the knowledge gained by a deep learning model can be transferred to other models. To achieve this deep pre-trained CNN models like VGG16, ResNet152 are available for transfer learning. Pre-trained models are the models that are trained on large amounts of data, and the weights updated during the training of the complex model can be applied to similar kind of tasks.

There are different types of pre-trained models which are trained on large scale datasets such as  ImageNet that consists of more than a million images. Popular pre-trained deep CNN models like VGG16, VGG19, ResNet152, InceptionV3, Xception, NASNet, Inception ResNet V2 and DarkNet are briefly described below: 

\begin{itemize}
    \item \textbf{Visual Geometric Group (VGG 16):} VGG16 is a deep ConvNet trained on 14 million images belonging to 1000 different classes and topped the leader board in ILSVR (ImageNet) challenge. In this architecture, 2X2 filters are used with stride 1 for convolution operation, and 2X2 filters with stride two and same padding are used for max-pooling operation across the network. At the end of architecture, two fully connected dense layers of 4096 neurons are connected followed by soft-max layer.
    \item \textbf{Neural Architecture Search Network (NASNet): } This is a special kind of Deep CNN which searches for a better architectural building block on small datasets like CIFAR10 and transfer it to larger datasets like ImageNet. It has a better regularisation mechanism called Scheduled drop path, which significantly improves generalisation. 
    
    \item \textbf{Xception: }
    Xception is another deep ConvNet architecture that supports depth-wise separable convolution operations and outperformed ResNet and InceptionV3 in  ILSVR challenge. 
    \item \textbf{Inception ResNetV2:} This is popularly known as InceptionV4, as it combines architectures of two different architectures called InceptionV3 and ResNet152. It has both inception and residual connections which boost the performance of the model. 
\end{itemize}

Deep neural networks give excellent performance only when trained with extensive data. If the data used to train is not sufficient, then the DNN models tend t overfit. Deep, Convolutional Neural Networks are introduced in \cite{simonyan2014very} for the task of Scalable Image Recognition. Xception, a deep CNN is developed using depthwise Separable convolutions to improve the performance \cite{chollet2017xception}. A flexible architecture has been defined in \cite{zoph2018learning}, which can search for a better convolutional cell with better regularisation mechanism. All these models are trained on ImageNet Dataset for ILSVR challenge.

Our objective is to create a robust and efficient model to recognise DR with limited datasets and with limited computational resources. To achieve our objective of creating a robust model with small datasets, we seek the help of transfer learning and use various pre-trained ConvNets to extract deep features.  We use the knowledge of these models to extract the most prominent features of colour fundus images. A deep neural network with dropout introduced at early layers is trained to detect and classify the severity levels of diabetic retinopathy. As we introduced dropout at the input layer, deep neural network is immune to over-fit.

\section{Proposed Methodology}
In this work, our objective is to develop a robust and efficient model to automate DR diagnosis. We focus on the extraction of deep features that are most descriptive and discriminate which ultimately improves the performance of DR recognition. In order to get an optimal representation, features are extracted from multiple pre-trained CNN architectures and are blended using pooling based approaches. These final representations are used to train a Deep Neural Network with a dropout at the input layer. Proposed model has three different modules: feature extraction, model training, and evaluation module. 
\begin{figure}[H]
    \centering
    \includegraphics[width=\linewidth]{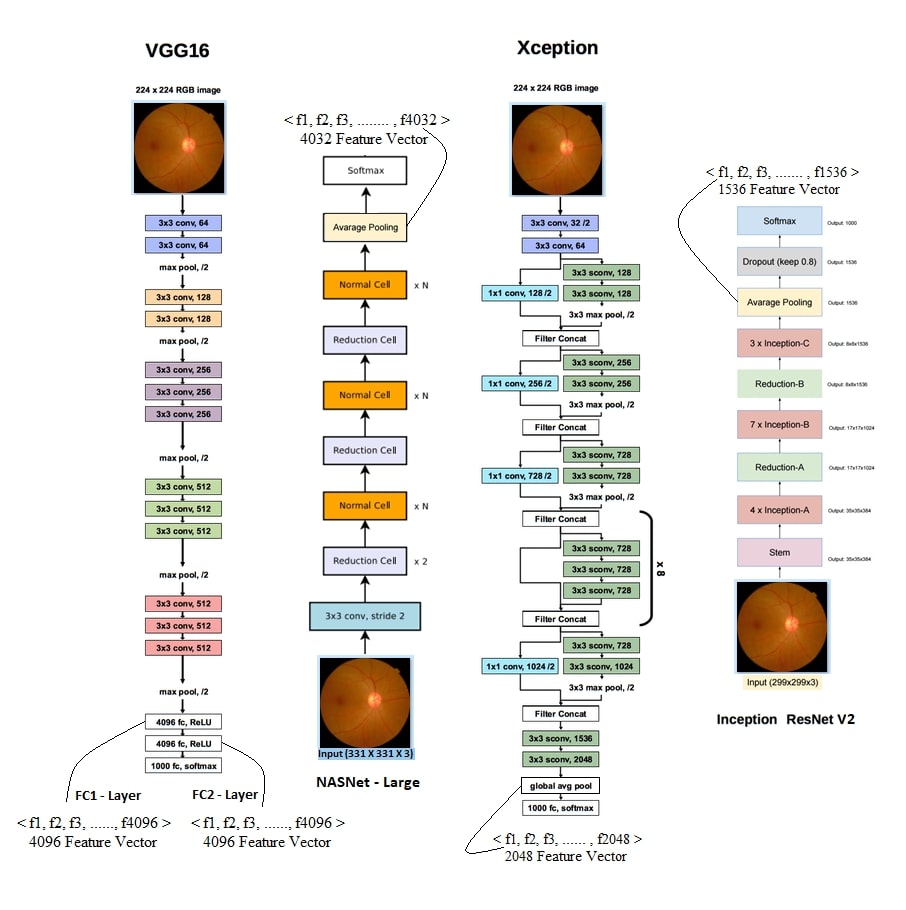}
    \caption{Architectures of various pre-trained models along with an indication of layers from which features are Extracted}
    \label{fig:pretrained_architectures}
\end{figure}

\subsection{Feature Extraction}
Performance of any machine learning model is highly influenced by the feature representations and the same is applicable to models used for DR recognition. With this motivation, we propose two different approaches (uni-modal and multi-modal) to extract optimal features from the given retinal images.

In the proposed work, initial representations of the retinal images are obtained from the pre-trained VGG16, NASNet, Xception Net and Inception ResNetV2. As each of the pre-trained model expects input  images of varying sizes, given retinal images are reshaped according to the input dimensions accepted by these models for example, when VGG16 is used images are reshaped to 224*224*3. These reshaped retinal images are fed to the pre-trained models after removing the soft-max layer and freezing the rest of the layers. 
Activation outputs from the penultimate layers form the basis for the proposed feature extraction module.
\noindent{For each  retinal image deep features are extracted from the pre-trained ConvNets and  following are the details:}
\begin{itemize}
\item Each of the first (fc1) and second (fc2) fully connected layers of VGG16 produces a feature vector of 4096 dimensions
\item The final global average pooling layer of NASNet, Xception and InceptionResNetV2 gives  feature vectors of size 4032, 2048 and 1536 respectively
\end{itemize}

Figure \ref{fig:pretrained_architectures} gives the architectural details of the pre-trained VGG16, NASNet, Xception and InceptionResNetV2 and pointers are marked at the feature extraction layers. These features form the input to the  proposed uni-modal and blended multi-modal approaches to obtain the optimal feature representations of the retinal images. 

\subsection{Uni-modal deep feature extraction: }
\begin{figure}[H]
    \center
    \includegraphics[width=\textwidth]{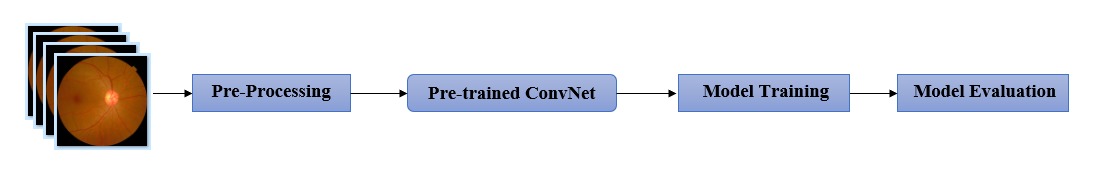}
    \caption{Stages involved in uni-modal deep feature based DR recognition}
    \label{fig:Unimodal}
\end{figure}

In this approach, deep features are extracted from the final layers of one of the pre-trained ConvNets (VGG16, NASNet, Xception, ResNet V2) to get the global representation of the retinal images. These deep features are fed to classification models for DR identification and recognition. 
We propose to use DNN architecture with a dropout at the input layer for DR identification and classification. Figure \ref{fig:Unimodal} gives the details of different stages involved in DR recognition process that uses uni-modal deep ConvNet features.

\subsection{Blended (multi-modal) deep feature extraction:}

Unlike uni-modal approaches,  multi-modal approaches use deep features extracted from multiple ConvNets and are blended using fusion techniques.
The features obtained from different pre-trained models provide a different representation of the retinal images as    they follow different architectures and are trained on different datasets. A stronger representation can be obtained by blending features from multiple ConvNets, as features of one ConvNet complements the features from other ConvNets involved in the process.

\begin{figure}[H]
    \center
    \includegraphics[width=\textwidth]{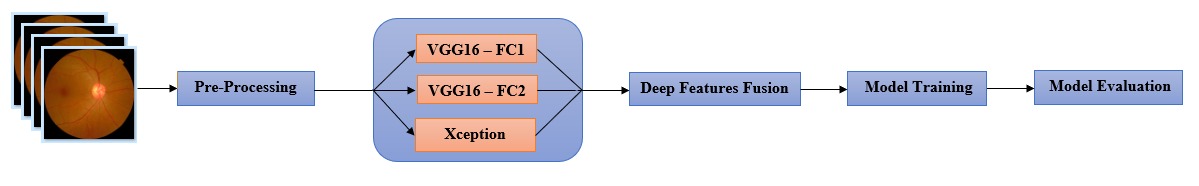}
    \caption{Stages involved in blended  deep feature based DR recognition}
    \label{fig:Multimodal}
\end{figure}
\begin{figure}[H]
    \center
    \includegraphics[width=\textwidth]{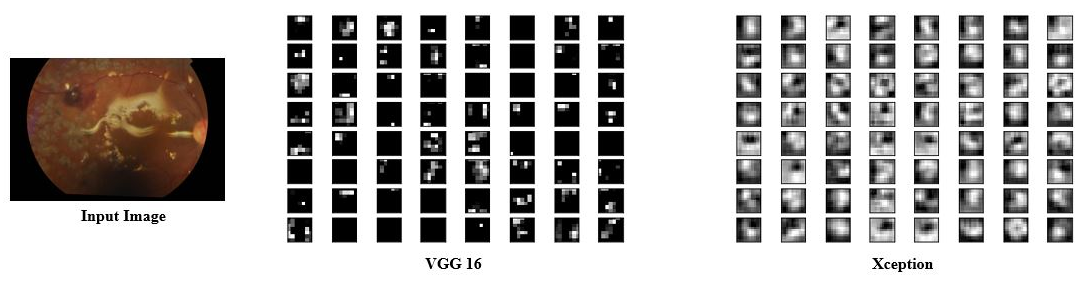}
    \caption{Visualization of the feature maps of the final convolution blocks of VGG16 and Xception models on passing retinal image as input}
    \label{fig:Convmaps}
\end{figure}

We propose various pooling approaches to fuse the deep features extracted from multiple pre-trained ConvNets. The final blended deep features provide better descriptive and discriminate representation of the retinal images. These blended features are fed to the classification models for DR identification or severity recognition. Figure \ref{fig:Multimodal} gives the details of different stages involved in DR recognition process that uses blended multi-modal deep ConvNet features.
The proposed blended multi-modal feature extraction module, uses features from both the fully connected layers of VGG16 (fc1 and fc2) and global average poling layer of Xception as input. 
The rationale behind choosing features VGG16 and Xception over others is two fold. 
In VGG16, each feature map of the final  convolution block learns the presence of different lesions from the retinal images.  Xception Net learns correlations across the 2-D space as a result each feature map provides  the comprehensive representation of the entire retinal scan. Figure \ref{fig:Convmaps} visualizes the feature maps obtained from the final convolution blocks of VGG16 and Xception models when a retinal image is passed to these models.

\subsubsection{Approaches to blend deep features from multiple ConvNets}
In this work, two different pooling based approaches (1-D pooling and cross pooling) are proposed to fuse multi-modal deep features that are extracted from VGG16 (fc1, fc2) and Xception. 1-D pooling is used to select  prominent local features from the each region of VGG16 whereas cross pooling allows to aggregate the prominent features obtained by 1-D pooling with global representation of Xception. 

1-D pooling based fusion takes one feature vector $U$ as input, and produces another feature vector $\hat U$, where $U \in R^{d1}$ , $\hat U \in R^{d2}$ and $d_2 \leq d_1 $. $\hat U$ is a reduced representation of $U$, where $U = \{u_1,u_2...u_{d1} \}$ and  $\hat U = \{\hat u_1,\hat u_2...\hat u_{d2} \}$. 
Each feature element $\hat u_i$, of the output vector $\hat U $ is computed using one of the following three approaches:
\begin{equation}
\text{1-D Max pooling:} \hat u_i = max(u_{i*2},u_{i*2 + 1}) \; \forall i \in \{1,2...d_2\}
\end{equation}
\begin{equation}
\text{1-D Min pooling: } \hat u_i = min(u_{i*2},u_{i*2 + 1}); \; \forall i \in \{1,2...d_2\}
\end{equation}
\begin{equation}
\text{1-D Average pooling: }  \hat u_i = mean(u_{i*2},u_{i*2 + 1}); \forall i \in \{1,2...d_2\}    
\end{equation}
\begin{equation}
\text{1-D Sum pooling:}  \hat u_i = u_{i*2} + u_{i*2 +1}; \forall i \in \{1,2...d_2\}
\end{equation}
In cross pooling based feature fusion, two different feature vectors X, Y are passed as input, and  another feature vector Z is produced, where ${X,Y,Z} \in R^{d}$. Each feature element $z_i$, of the output vector $Z$ is computed using one of the following three approaches:
\begin{equation}
\text{Cross Max pooling: } z_i = max(x_i,y_i) \quad \forall i \in \{1,2...d\}
\end{equation}
\begin{equation}
\text{Cross Min pooling:} z_i = min(x_i,y_i) \quad \forall i \in \{1,2...d\}
\end{equation}
\begin{equation}
\text{Cross Average pooling:} z_i = mean(x_i,y_{i+1}) \forall i \in \{1,2...d\}
\end{equation}
\begin{equation}
\text{Cross Sum pooling:}\quad  y_i = x_{i} + y_{i}\qquad \forall i \in \{1,2...d\}
\end{equation}

\begin{figure}[h]
    \centering
    \includegraphics[width=\textwidth]{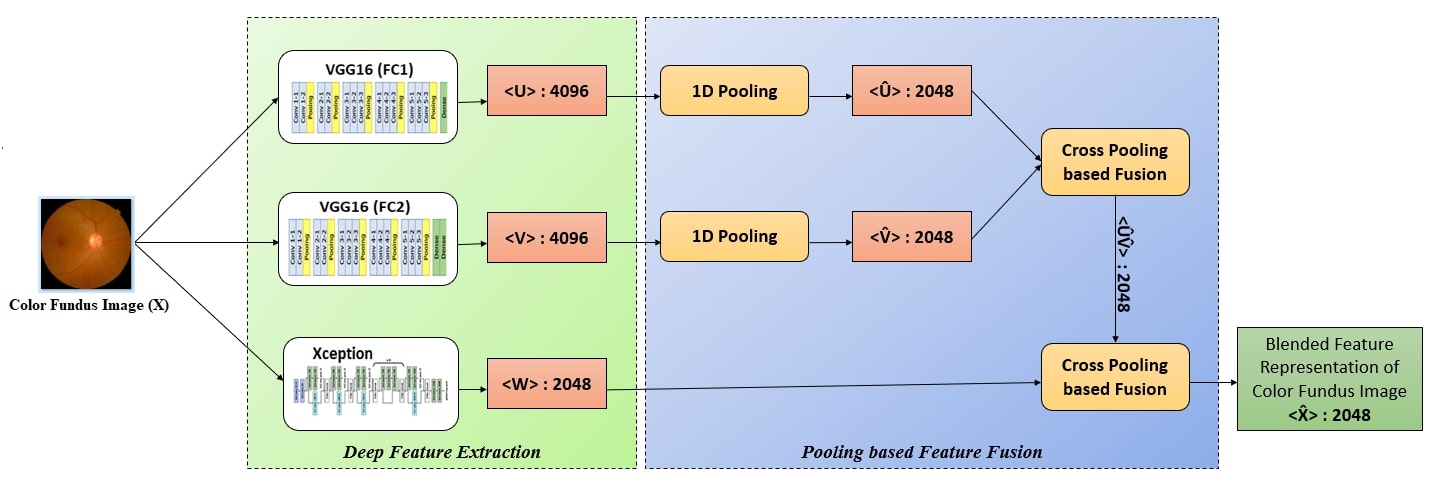}
    \caption{Approaches for Fusion of features extracted from Deep ConvNets}
    \label{fig:proposed}
\end{figure}

1-D pooling is applied independently on features extracted from fc1 and fc2 layers of VGG16. Then cross pooling approach is applied on the resultant pooled features. This feature vector is merged with the features extracted from the Xception using cross pooling. Fusion module produces deep blended features, which are used to train the proposed DNN model.  Figure \ref{fig:proposed} shows the proposed architecture of the deep feature fusion approach used to blend features from different ConvNets. As the final feature vector is a blended version of the local and global representations of the retinal images it provides strong features. Algorithm \ref{algo} gives the sequence of steps involved in the blended multi-modal feature fusion based DR recognition.  

\begin{algorithm*}
    \SetAlgoLined
    \KwIn{Let $D^{Tr}$ and $D^{Tst}$ be the train and test datasets of fundus images respectively where  $D^{Tr} = \{(x_i,y_i)_{i=1}^{N_{Tr}}\}$ and $D^{Tst} = \{(x_i)_{i=1}^{N_{Tst}}\}$. ${x_i}$ represents $i^{th}$ color fundus image in the dataset and $y_i$ is the severity level of DR associated with $x_i$. In case of DR identification task $y_i \in \{0,1\}$ whereas in the case of DR severity classification task $y_i \in \{0,1,2,3\}$.}
    \KwOut{$y_i$ for each $x_i \in D^{Tst}$}   
    \textbf{Step1:} Preprocess each image $x_i$ in the dataset. \\ 
    \textbf{Step2:} Feature Extraction \\
    $\quad$ For each preprcossed image $x_i$ three different features ($V_i, U_i, W_i$) are extracted. \\
    $ \qquad V_i \leftarrow $ Features extracted from fc1 layer of VGG16\\
    $ \qquad U_i \leftarrow $ Features extracted from fc2 layer of VGG16\\
    \qquad $W_i \leftarrow $ Features extracted from global avg pool layer of Xception\\
    \quad Where $V_i \in d1, U-i \in d2$ and $W_i \in d3$ \\ 
    \textbf{Step3:} Deep feature fusion\\
    $\quad$ Apply feature feature fusion on the deep features extracted from each image\\
    $\qquad \hat V_i \leftarrow  max(V_{i*2},V_{i*2 + 1)}; \quad \forall i \in \{1,2...d_1\} \qquad $  (1-D max pooling)\\
    \qquad $\hat U_i \leftarrow  max(U_{i*2},U_{i*2 + 1)}; \quad \forall i \in \{1,2...d_2\} \qquad $  (1-D max pooling)\\
    \qquad $\hat{UV}_i \leftarrow \frac{ (\hat V_{i}+ \hat U_{i})}{2}; \quad \forall i \in \{1,2...d_2\} \qquad $  (Average Cross pooling)\\
    \qquad $\hat{x}_i \leftarrow \frac{ (\hat {\hat{UV}}_{i}+  W_{i})}{2}; \quad \forall i \in \{1,2...d_3\} \qquad $  (Average Cross pooling)\\
    \qquad $\hat x_i:$ blended feature vector corresponding to $x_i$\\ 
    \textbf{Step4:} Model Training\\
    \quad Training dataset  is prepared using the blended features $D^{Tr} = \{(\hat x_i,y_i)\}_{i=1}^{N_{Tr}}$ \\
    \quad Train a deep neural network (DNN) using $D^{Tr}$\\ 
    \textbf{Step5:} Model evaluation\\
    \quad Test dataset is prepared using the blended features $D^{Tst} = \{(\hat x_i)\}_{i=1}^{N_{Tst}}$ \\
    \quad Evaluate the performance of $D^{Tst}$using the DNN trained in \textbf{Step4}\\
    \caption{Blended multi-modal deep feature fusion based DR recognition task}
    \label{algo}
\end{algorithm*}

\subsection{Model Training and Evaluation:}
During this phase, we train the ML model with deep blended pre-trained features. We prefer to use Deep Neural Network (DNN) model for training. For DR identification task, as it is a simple binary classification task, a DNN with two hidden layers with 256, 128 units respectively with ReLU activation is used. 

\begin{figure}[H]
    \center
    \includegraphics[width=\textwidth]{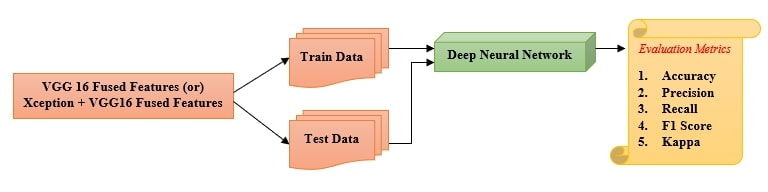}
    \caption{Training and Evaluation of DNN model for identification and recognition of DR}
    \label{fig:model}
\end{figure}

For DR severity classification task, a DNN with three hidden layers with 512, 256, 128  units respectively using ReLU activation is used. 
For both the DNNs with the input layer we applied 0.2 dropout to avoid model from over-fitting of model. This helped the model to become robust.  Figure \ref{fig:model} represents the architecture of proposed approach for model training and evaluation. 

\section{Experimental Results}
In this section, we provide details of experimental studies that are being carried out to understand the efficiency of the proposed blended multi-modal deep features representation. \subsection{Dataset Summary}
For the experimental studies, the APTOS 2019 kaggle benchmark dataset available as part of the blindness detection challenge is used \cite{APTOS}.
This is a large dataset of retinal images taken using fundus photography under a variety of imaging conditions. The images are graded manually on a scale of 0 to 4 (0 - No DR, 1-Mild, 2-Moderate, 3-Severe, 4-Proliferative DR) to indicate different  severity levels.
\begin{table}[htbp]
    \begin{center}        \renewcommand{\arraystretch}{1.5}                \begin{tabular}{|l|c|}
            \hline
            \textbf{Severity Level} & \textbf{\# Samples} \\ \hline
            \textbf{Class-0 (Normal)} & 1805 \\ \hline
            \textbf{Class-1 (Mild Stage)} & 370\\ \hline
            \textbf{Class-2 (Moderate Stage)} & 999 \\ \hline
            \textbf{Class-3 (Severe Stage)} & 193\\ \hline
            \textbf{Class-4 (Proliferative Stage)} & 295\\ \hline
            \textbf{Total} & 3662\\ \hline
        \end{tabular}
        \caption{Dataset Summary of APTOS 2019 dataset} \label{tab:severity_dataset}
    \end{center}
\end{table}
\vspace{-2em}
Table \ref{tab:severity_dataset} gives the number of retinal images available in the dataset under each level of severity. We can observe that the dataset has an imbalance with more number of normal images, and with very few images in class3.
In all the experiments,  80\% of the data is used for training and the remaining 20\% is used for validation.

\subsection{Performance Measures:}
For the evaluation of the proposed model, we report different measures: Accuracy, Precision, Recall, and F1 Score. In addition, we used an additional metric called Kappa statistic to compares an observed accuracy with an expected accuracy.
\noindent{Kappa Statistic is calculated as}
\[\text{Kappa Score} = \frac{(\text{Observed Accuracy} -\text{Expected Accuracy})}{(1-\text{Expected Accuracy})}\]
\noindent{Observed accuracy is defined as the number of samples that are correctly classified. Expected accuracy is defined as the accuracy that a classifier would be expected to achieve, which is directly related to the number of examples of each class, along with the number of examples that the predicted value satisfied with the correct label.}
\subsection{DR Identification and Severity level Prediction:}
The whole set of experiments carried out in this work are divided into two different tasks. In task1, presence or absence of DR is identified where as in task2,   severity level is predicted for the given retinal image.
\subsubsection{Task1 - DR Identification:}
In this task, given the DR image of a diabetic patient, we need to check whether the person is effected by retinopathy or not. DR identification is a binary classification task, so binary cross entropy loss is used to measure the loss, and Adam optimiser is used to optimise the objective function.
The dataset contains images belonging to 5 different  classes as shown in table \ref{tab:severity_dataset} and is not suitable for binary  classification task. Merging all the DR effected images into a single class gives 1857  positively labeled images and the remaining 1805 normal images are   labeled as negative.

\subsubsection{Task2 - Severity level Prediction:}
Objective of task1 is to identify the presence or absence of DR, given a retinal image. While treating the DR effected patients, mere identification of DR would not be sufficient and understanding the level of severity would be helpful for better treatment. Hence we treat severity level identification as a  separate task that  categorises the given retinal image to one of the 5 severity levels. Categorical Cross entropy loss is used to represent loss and Adam optimiser is used to optimise the objective function. 
\subsection{Experimental studies to show the representative nature of uni-modal features for task1}
This experiment is carried out to understand how  efficiently retinal images are represented using uni-modal features that are directly obtained from single pre-trained ConvNet.
Models like 
VGG16, Xception, NASNET, and ResNetV2 are considered to extract uni-modal features. 
For classification, models like Naïve Bayes classifier, logistic regression, decision tree, k-Nearest Neighbourhood (KNN) classifier, Multi Layered Perceptron (MLP) Support Vector Machine (SVM) and Deep Neural Network (DNN) are used.

\begin{table}[h!]
    \centering
    \caption{Performance of ML algorithms on Task1 using features from fc2 layer of VGG16}
    \label{tab:result_binary_vgg16_fc2}
    \begin{tabular}{|c|c|c|c|c|c|}
        \hline
        \textbf{Model} & \textbf{Accuracy} & \textbf{Precision} & \textbf{Recall} & \textbf{F1 Score} & \textbf{Kappa Statistic} \\ \hline
        Logistic Regression & 97.13 & 97 & 97 & 97 & 94.27 \\ \hline
        KNN                 & 95.36 & 96 & 95 & 95 & 90.73 \\ \hline
        Naive Bayes         & 77.08 & 82 & 77 & 76 & 54.45 \\ \hline
        Decision Tree       & 91.27 & 91 & 91 & 91 & 82.52 \\ \hline
        MLP                 & 96.45 & 97 & 96 & 96 & 92.91 \\ \hline
        SVM (linear)        & 96.58 & 97 & 97 & 97 & 93.17 \\ \hline
        SVM (RBF)           & 96.86 & 97 & 97 & 97 & 93.73 \\ \hline
        DNN                 & 97.32 & 98 & 98 & 98 & 94.63 \\ \hline
    \end{tabular}
\end{table}

Table \ref{tab:result_binary_vgg16_fc2} and \ref{tab:result_binary_xception}  shows the performance of DR identification task using different ML models  when the retinal images are represented with the features extracted from the first fully connected layer (fc2) of VGG16 and Xception respectively. 
With this we came to a conclusion that DNN outperforms the rest of the ML model irrespective of the models. Hence decided to use DNN model alone in the rest of the experiments.

1SW \begin{table}[h!]
    \centering
    \caption{Performance of ML algorithms on Task1 using features from Xception}
    \label{tab:result_binary_xception}
    \begin{tabular}{|c|c|c|c|c|c|}
        \hline
        \textbf{Model} & \textbf{Accuracy} & \textbf{Precision} & \textbf{Recall} & \textbf{F1 Score} & \textbf{Kappa Statistic} \\ \hline
        Logistic Regression & 96.45 & 96 & 96 & 96 & 93    \\ \hline
        KNN                 & 95.5  & 96 & 95 & 95 & 91    \\ \hline
        Naive Bayes         & 82.95 & 84 & 83 & 83 & 65.9  \\ \hline
        Decision Tree       & 87.59 & 88 & 88 & 88 & 75.17 \\ \hline
        MLP                 & 96    & 96 & 96 & 96 & 91.89 \\ \hline
        SVM (linear)        & 96.18 & 96 & 96 & 96 & 92.36 \\ \hline
        SVM (RBF)           & 97.4  & 97 & 97 & 97 & 94.82 \\ \hline
        DNN                 & 97.41 & 97 & 97 & 97 & 94.82 \\ \hline
    \end{tabular}
\end{table}

\begin{table}[h!]
    \centering
    \caption{Task1 performance using DNN  trained on different uni-modal features}
    \label{tab:results_task1_uni}
    \begin{tabular}{|c|c|c|c|c|c|}
        \hline
        \textbf{Model} & \textbf{Accuracy} & \textbf{Precision} & \textbf{Recall} & \textbf{F1 Score} & \textbf{Kappa Statistic} \\
        \hline
VGG16-fc1                  & 97.27 & 97 & \textbf{98} & 97 & \textbf{95.12} \\ \hline
VGG16-fc2 & 97.32 & \textbf{98} & \textbf{98} & 98 & 94.63 \\ \hline
                NASNet & 97.14 & 97 & 97 & 97 & 94.27 \\ \hline
    Xception & \textbf{97.41} & 97 & 97 & 97 & 94.82 \\ \hline
Inception ResNetV2                & 97.34 & 97 & 97 & 97 & 94.54 \\ \hline
    \end{tabular}
\end{table}

Table \ref{tab:results_task1_uni} shows the representative power of uni-modal features that are extracted from different pre-trained models.
It is clear from the results that the performance of the DNN model varies depending on the uni-modal features used. This experiment gives a clue that each pre-trained model extracts a different set of features from retinal images. 
The features extracted from Xception yields better performance in terms of accuracy for the diabetic retinopathy identification task. A nominal difference in terms of accuracy and kappa score can be observed between the models trained using different uni-modal features.

\begin{table}[htbp]
    \begin{center}  \caption{Task1-Comparison of DNN model (trained on uni-modal features) in terms of loss and number of epochs when trained on different uni-modal features } \renewcommand{\arraystretch}{1.5}    \begin{tabular}{|l|c|c|c|} \hline    \textbf{Model} & \textbf{\# epochs} & \textbf{loss} & \textbf{Accuracy}  \\ \hline
            VGG16-fc1 & 65 & 0.0024 & 97.27 \\ \hline
            VGG16-fc2 &    67 & 0.0139 & 97.32  \\ \hline
            NASNet    & 37 & 0.0310 & 97.14 \\ \hline
            Xception & 16 & 0.0213 & 97.41 \\ \hline
            Inception ResNet V2    & 19 & 0.0815 & 97.34 \\ \hline        \end{tabular}     \label{tab:result_binary_epochs}
    \end{center}
\end{table}

For a better understanding of the representative nature of different uni-modal features, loss and number of epochs taken to converge by the DNN models are reported in  Table \ref{tab:result_binary_epochs}.  We can observe that the model trained using VGG16-Fc1 reaches minimum loss compared to the rest of the models. In terms of convergence, Xception takes only 16 epochs whereas performance of Inception ResNetV2 outperformed other models.

To summarize the experiments on DR identification task, features extracted from Xception, VGG16-fc2 and Inception ResnetV2 yields  the same accuracy with nominal differences. However, models trained on the VGG16-fc1 features gives better kappa scores compared to others. We can also observe that models trained on the VGG16-fc2 features gives better performance in terms of  precision, recall and F1 scores. Regardless of the type of uni-modal features used, DNN consistently outperforms rest of the models especially in terms of kappa scores.
The reason for the superior performance of the models trained using VGG16 and Xception features is that these models are good at extracting the lesion information that is useful to discriminate the DR effected images  from those that are not effected.
\subsection{Experimental studies to show the representative nature of uni-modal features for task2} We run a set of experiments to understand the nature of uni-modal features for  severity prediction of DR. Task2 is more challenging compared to task1 as it involves multiple  classes. DNN model with dropout at the  input layer is used with different uni-modal features. 
\begin{table}[h!]
    \centering
    \caption{Task2 performance using DNN  trained on different uni-modal features}
    \label{tab:result_multi_vgg16_fc1}
    \begin{tabular}{|c|c|c|c|c|c|}
        \hline
        \textbf{Type of  Uni-modal features} & \textbf{Accuracy} & \textbf{Precision} & \textbf{Recall} & \textbf{F1 Score} & \textbf{Kappa Statistic} \\ \hline
        VGG16-fc1 & \textbf{80.06} & \textbf{80} & \textbf{81} & \textbf{80} & \textbf{70.02} \\
        \hline
        VGG16-fc2  & 79.81 & 79 & 80 & 79 & 68.88 \\ \hline
NASNET & 76.4  & 75 & 76 & 75 & 63.87 \\ \hline
Xception            & 78.99 & 78 & 79 & 78 & 67.67 \\ \hline
Inception ResNetV2  & \textbf{79.73} & 78 & 78 & 78 & 67.67 \\ \hline
    \end{tabular}
\end{table}

Based on the results reported in Table \ref{tab:result_multi_vgg16_fc1}, we can observe the same trend that has been observed in task1. The scores obtained for task2 shows the complexity of severity prediction. The model trained on VGG-16+fc1 features shows superior performance than rest of the models. The same can be observed in terms of all the metrics. 

\begin{table}[htbp]
    \begin{center}
\caption{Task2-Comparison of DNN model (trained on uni-modal features) in terms of loss and number ofepochs when trained on different uni-modal features }    \renewcommand{\arraystretch}{1.5}        \small        \begin{tabular}{|l|c|c|c|} \hline        \textbf{Model} & \textbf{\# epochs} & \textbf{loss} & \textbf{Accuracy}  \\ \hline
            VGG16-fc1 & 76 & 0.3623 & 80.06 \\ \hline
            VGG16-fc2 &    79 & 0.3986 & 79.81  \\ \hline
            NASNet    & 37 & 0.5612 & 76.39 \\ \hline
            Xception & 23 & 0.4175 & 78.99 \\ \hline
            Inception ResNet V2    & 89 & 0.382 & 79.73 \\ \hline    \end{tabular}      \label{tab:result_task2_uni_epochs}
    \end{center}
\end{table}
From Table \ref{tab:result_task2_uni_epochs} it is clear that  among all the pre-trained features, VGG16-fc1 yields superior performance with minimum loss. However Xception converges in lesser number of epochs compared to other models.
\subsection{Performance evaluation of the proposed blended multi-modal features}
A clue from the experiments on uni-modal features is that different uni-modal features extract different sets of features from the retinal images.  If we can use multiple deep features extracted from different models, they complement each other and helps to improve the scores.
To get benefited from more than one set of uni-modal features we propose a blended multi-modal feature representation. This section is dedicated to show the representative power of the proposed feature representation with an application to DR identification and severity level prediction.

In addition we apply the proposed pooling methods to blend the features from multiple pre-trained models. Initially we blend features from first and second fully connected layers of VGG16. Then we extend this to fusion of 3 different features from fc1, fc2 layers of VGG16 and Xception.

\subsubsection{Blended Multi-Modal  deep features for task1} 

\begin{table}[h!]
    \centering
    \caption{DNN with blended  multi-modal features with different fusions for Task1}
    \label{tab:result_task1_multi}
    \begin{tabular}{|l|l|c|c|c|c|}
        \hline Modalities &
        Pooling & Accuracy & Kappa Statistic & Epochs & Loss   \\ \hline    \multirow{3}{*} {VGG16-fc1  and  VGG16-fc2} & Max-pooling  & 96.12    & 91.89           & 68     & 0.0352 \\ \cline{2-6}
        & Avg-pooling  & \textbf{97.39 }   & \textbf{94.61 }          & \textbf{51}     & \textbf{0.0293} \\ \cline{2-6}
        & Sum-pooling      & 95.5     & 91             & 64     & 0.0419 \\ \hline
        \multirow{3}{*} {
        VGG16-fc1, VGG16-fc2 and Xception} &    Max-pooling  & 96.85    & 92.6            & 69     & 0.0314 \\ \cline{2-6} &
        Avg-pooling  & \textbf{97.92}    & \textbf{94.93 }          &\textbf{ 43 }    & \textbf{0.0201} \\ \cline{2-6} & Sum-pooling      & 96.1     & 92.31           & 56     & 0.0396 \\ \hline
        
    \end{tabular}
\end{table}

We experiment the effect of   blending deep features extracted from multiple  pre-trained models on DR identification task.
In addition we verify the proposed maximum, sum and average pooling approaches to blend multiple deep features. 

From Tables \ref{tab:result_task1_multi}, we can observe that  average pooling  based fusion works better for DR Detection compared to other models.  
Using average fusion the models trained on multi-modal features leads to superior performance in terms of accuracy and kappa static. In addition the model converges faster in less than 50 epochs and attains minimum loss. The accuracy obtained by model trained using multi-modal features is significantly better compared with to those trained on uni-modal features.

\subsubsection{Blended Multi-Modal deep features for task2}
From the previous experiments we understand that the models trained on multi-modal features give better performance compared to those trained on uni-modal features in the context of DR identification which is simple binary task.
To understand that the proposed blended performs efficiently for more complex  multi-class classification task, we apply the proposed feature representation for severity prediction task. 

\begin{table}[h!]
    \centering
    \caption{DNN with blended  multi-modal features with different fusions for Task2}
    \label{tab:result_binary_vgg16pool}
    \begin{tabular}{|l|l|c|c|c|c|}
        \hline Modalities &
        Pooling & Accuracy & Kappa Statistic & Epochs & Loss   \\ \hline    \multirow{3}{*} {VGG16-fc1  and  VGG16-fc2} & Maximum  & 78.06    & 66.87    & 72  
        & 0.4176 \\ \cline{2-6}
        &         Average  & \textbf{80.34}    & \textbf{69.21 }     &\textbf{ 62 }    & \textbf{0.2987 } \\ \cline{2-6}
        &         Sum      & 76.8     & 65.64           & 68     & 0.5693 \\ \hline
        \multirow{3}{*} {
        VGG16-fc1, VGG16-fc2 and Xception} & Maximum  & 79.25    & 67.29           & 74     & 0.3986 \\  \cline{2-6} &          Average  & 80.96    & 70.9            & 54     & 0.2619 \\ \cline{2-6}
        & Sum      & 77.12    & 66.42           & 61     & 0.4782 \\ \hline
    \end{tabular}
\end{table}
From Table  \ref{tab:result_multi_xception_vgg16pool}, we can see that average pooling based fusion of multiple deep features works better for Diabetic Severity Prediction. Compared to the blended features   from VGG16-fc1 and VGG16-fc2, blended features from VGG16-fc1, VGG16-fc2 and xception gives  better representation.
For severity prediction also, the model that uses average pooling  approach for fusion converges faster with better accuracy and kappa score when compared with other approaches for fusion.
\begin{figure}[H]
 \centering
 \includegraphics[width=.5\linewidth]{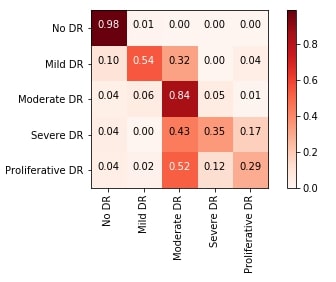}
 \caption{ {Confusion matrix for the severity prediction task.}}
 \label{fig:confmat}
\end{figure}
\subsection{Comparison of proposed Blended feature extraction with existing methods}
In this experiment we show the effectiveness of  the proposed DNN with dropout at the input layer trained using the proposed blended multi-modal deep feature representation.   with the existing models in the literature for DR prediction. We compare the proposed model with the performances of the models used in \cite{gargeya2017automated} and \cite{kassani2019diabetic}. From Table  \ref{tab:result_multi_xception_vgg16pool} we can see that the proposed method gives an  accuracy of 80.96\% which is significantly better than existing models in the literature. 
When compared to the existing models proposed DNN model is simple with only 3 hidden layers with 512, 256, and 128 units each hidden layer. 
Confusion matrix in Figure \ref{fig:confmat}  shows the mis-classifications produced by the proposed model when applied for DR severity prediction task. From the figure we can see that most of the proliferate DR type images are  predicted as moderate.    

As the final feature vector is a blended version of the local and global representations of the retinal images the final representation provides strong features. The reason for improvement in the performance of the proposed model is that each feature map of the final convolution block of VGG16 learns the presence of different lesions from the retinal images and  Xception Net  comprehensive representation of the entire retinal scan. When we combine the deep features from VGG16 and Xception gives a compact representation that gives the wholistic representation of DR images.  

\begin{table}[H]
    \begin{center}\caption{Comparison of Proposed method using with existing methods} \renewcommand{\arraystretch}{1.5}        \begin{tabular}{|l|c|} \hline
            \textbf{Model} &  \textbf{ Accuracy} 
            \\ \hline
            DR detection using Deep Learning \cite{gargeya2017automated} & 57.2\% \\ \hline
            DR Classification Using Xception \cite{kassani2019diabetic} & 79.59      \\ \hline
            DR Classification Using InceptionV3 \cite{kassani2019diabetic}& 78.72       \\ \hline
            DR Classification Using MobileNet \cite{kassani2019diabetic}& 79.01      \\ \hline
            DR Classification Using ResNet50 \cite{kassani2019diabetic}& 74.64     \\ \hline
            Blended features + DNN (proposed) & \textbf{80.96}     \\ \hline
        \end{tabular}    
        \label{tab:result_multi_xception_vgg16pool}
    \end{center}
\end{table}

\section{Conclusion}
Major objective of this work is to acquire a compact and comprehensive representation of retinal images as the feature representations extracted from retinal images significantly  influence the performance of DR prediction.    
Initially we extract features from deep pre-trained VGG16-fc1, CGG16-fc2 and Xception models. VGG16 model learns the lesions and Xception learns the global representation of the images. Then the features from multiple ConvNets are blended to get final prominent representation of colour fundus images. The final representation is a obtained by pooling the representations from VGG16 and Xception features. A DNN model trained using these blended features for the task of Diabetic Retinopathy severity level prediction. The proposed DNN model with dropout at the input avoids over-fitting and converges faster. Our  experiments on benchmark APTOS 2019 dataset shows the superiority of the proposed model when compared to the existing models. 
Among the proposed pooling approaches, average pooling used to fuse the features extracted from the penultimate layers of multiple pre-trained ConvNets gives better performance with minimum loss in fewer epochs compared to others. 

\vspace{6pt}
\authorcontributions{Conceptualization, J.D.B.; methodology, S.N.S; software, S.H.; validation, M.B., P.K.R.M.
and O.J.; formal analysis, O.J.; investigation, P.K.R.M.; resources, J.D.B.; writing–original draft preparation, J.D.B.;
writing–review and editing, N.V.; visualization, S.N.S.; supervision, M.B.; project administration, S.H.; funding acquisition, O.J. All authors have read and agreed to the published version of the manuscript.}

\funding{This work was supported by the National Research Foundation of Korea (NRF) grant funded by the Korea government (MSIT) under Grant NRF-2018R1C1B5045013.}

\conflictsofinterest{The authors declare no conflict of interest.}

\vspace{6pt} 

\reftitle{References}

\end{document}